# Interface induced Zeeman-protected superconductivity in ultrathin crystalline lead films

Yi Liu[1,2,†], Ziqiao Wang[1,2,†], Xuefeng Zhang[1,2], Chaofei Liu[1,2], Yongjie Liu[3], Zhimou Zhou[1,2], Junfeng Wang[3], Qingyan Wang[1,2,4], Yanzhao Liu[1,2], Chuanying Xi[5], Mingliang Tian[5], Haiwen Liu[6,*], Ji Feng[1,2], X. C. Xie[1,2] and Jian Wang[1,2,7,*]

[1]*International Center for Quantum Materials, School of Physics, Peking University, Beijing 100871, China.*
[2]*Collaborative Innovation Center of Quantum Matter, Beijing 100871, China.*
[3]*Wuhan National High Magnetic Field Center, Huazhong University of Science and Technology, Wuhan 430074, China.*
[4]*Institute of Physics, Chinese Academy of Sciences, Beijing 100190, China.*
[5]*High Magnetic Field Laboratory, Chinese Academy of Sciences, Hefei 230031, Anhui, China.*
[6]*Center for Advanced Quantum Studies, Department of Physics, Beijing Normal University, Beijing 100875, China.*
[7]*CAS Center for Excellence in Topological Quantum Computation, University of Chinese Academy of Sciences, Beijing 100190, China*

†These authors contributed equally to this work.
*Corresponding authors: Haiwen Liu (haiwen.liu@bnu.edu.cn) and Jian Wang (jianwangphysics@pku.edu.cn)

**Two dimensional (2D) superconducting systems are of great importance to exploring exotic quantum physics. Recent development of fabrication techniques stimulates the studies of high quality single crystalline 2D superconductors, where intrinsic properties give rise to unprecedented physical phenomena. Here we report the observation of Zeeman-type spin-orbit interaction protected superconductivity (Zeeman-protected superconductivity) in 4 monolayer (ML) to 6 ML crystalline Pb films grown on striped incommensurate (SIC) Pb layers on Si(111) substrates by molecular beam epitaxy (MBE). Anomalous large in-plane critical field far beyond the Pauli limit is detected, which can be attributed to the Zeeman-protected superconductivity due to the in-plane inversion symmetry breaking at the interface. Our work demonstrates that in superconducting heterostructures the interface can induce Zeeman-type spin-orbit interaction (SOI) and modulate the superconductivity.**

2D superconducting system has raised considerable interests in condensed matter physics. The development of state-of-art film growth technique makes it possible to investigate the intrinsic properties of high quality crystalline superconducting films down to a few atomic layers, such as Pb and In films on Si [1-8], Ga films on GaN [9,10] and FeSe films on SrTiO$_3$ [11,12]. In general, the in-plane critical field $B_c$ of ultrathin superconducting films mainly depends on the spin pair breaking effect, namely the Pauli limit $B_p$ [13,14]. However, several mechanisms can weaken or eliminate the spin pair breaking effect, and enhance $B_c$ beyond the Pauli limit. For instance, in spin-triplet superconductors, the Cooper pairs formed by two electrons with parallel spin orientation are immune to the spin pair breaking effect and hence can easily overcome the Pauli limit [15]. In disordered 2D systems with conventional superconducting pairing, Pauli limit can be surpassed by spin-orbit scattering, which randomizes the spin orientation and thus weakens spin paramagnetism [16,17]. Besides, intrinsic SOI can also give rise to the enhancement of $B_c$. Rashba-type SOI, which generates from the out-of-plane inversion asymmetry, can polarize the spin of the electron to the in-plane direction, but the in-plane critical field $B_c$ cannot exceed $\sqrt{2}B_p$ [18]. In highly crystalline 2D superconductors with in-plane inversion symmetry breaking, such as single layer NbSe$_2$ and MoS$_2$, extremely large in-plane critical field far beyond the Pauli limit is observed and attributed to Ising superconductivity [19-23]. In these systems, the broken in-plane inversion symmetry gives rise to Zeeman-type SOI, causing an out-of-plane spin polarization of energy valleys, and hence protects the superconductivity under high in-plane magnetic field. To fully understand the general effect of Zeeman-type SOI on the in-plane critical field, more experimental and theoretical studies on various 2D crystalline superconductors beyond TMDs are necessary. Nevertheless, most 2D superconductors naturally preserve in-plane inversion symmetry, where Zeeman-protected superconductivity cannot exist.

Here we report our observation of interface induced Zeeman-protected superconductivity in ultrathin crystalline Pb films epitaxially grown on SIC phase on Si(111) substrate. Under parallel magnetic field, superconductivity survives at high fields far beyond the Pauli limit, suggesting Zeeman-protected superconductivity in ultrathin Pb films. The temperature dependence of the in-plane critical field qualitatively deviates from Klemm-Luther-Beasely theory of spin-orbit impurity scattering mechanism [16]. Furthermore, *ab initio* calculations show that the special lattice distortion in SIC phase can extend to the Pb films, which is very likely to induce Zeeman-type SOI and results in the pronounced spin splitting of the bandstructure in Pb film/SIC system.. Based on the microscopic analysis, we find that the Zeeman-type SOI largely enhances the in-plane critical field, and is quantitatively consistent with our experimental observation.

The atomically flat ultrathin crystalline Pb films were epitaxially grown on a Pb-induced SIC layer on Si(111) substrate in an ultrahigh-vacuum MBE chamber. The SIC layer has a thickness of 0.26 nm and is acquired by deposition of 1.5 ML Pb on Si(111) substrate at room temperature and subsequent annealing at about 573 K for 30 seconds. A typical scanning tunneling microscopy (STM) image of 4 ML Pb film is shown in Fig. 1. For *ex situ* transport measurements, Pb films with a nominal thickness of 4, 5, 6 were protected and immobilized by 6 nm amorphous Si capping layer. The substrate and the protection layer become insulating below 160 K, providing an ideal platform to study the superconducting properties of Pb films.

Figure 2 summarizes the transport results of 4 ML Pb film down to 0.5 K. The schematic of the transport measurement is presented in the inset of Fig. 2(a). According to temperature dependent sheet resistance $R_{sq}$ curve at zero magnetic field, the superconducting transition occurs at $T_c^{onset} = 6.23$ K indicated by extrapolating both the normal state resistance and the superconducting transition curve [Fig. 2(a)], and the resistance drops to zero within the measurement resolution at $T_c^{zero} = 5.78$ K. The superconductivity in 4 ML Pb film is gradually suppressed by increasing perpendicular magnetic fields [Fig. 2(b) and 2(c)]. Moreover, the upper critical field of 4-6 ML Pb films is plotted in Fig. S2, which is well fitted by Werthamer-Helfand-Hohenberg (WHH) theory [24], yielding a larger critical field for thinner film. Fig. 2(d) shows the $V(I)$ characteristics measured at various temperatures from 5.8 K to 6.5 K in the absence of magnetic field. A power-law dependence of $V \sim I^\alpha$ is observed, where $\alpha$ is the slope of $V(I)$ curves plotted in double-logarithmic scale. The exponent $\alpha$ reduces with increasing temperature and approaches 3 at 6.10 K (defined as $T_{BKT}$, see the inset of Fig. 2(d)), indicating a Berezinski-Kosterlitz-Thouless (BKT)-like behavior [19,25,26] in 4 ML Pb film.

Figure 3 shows the main results of the transport measurements on the 4-6 ML Pb films under parallel magnetic field. $R_{sq}(B)$ curves measured on 4, 5 ML Pb films up to 15 T are displayed in Fig. 3(a), 3(b), respectively. A large in-plane critical field (defined as the magnetic field required to reach 50% of the normal state resistance $R_n$) is observed and extrapolated beyond the Pauli limit [the inset of Fig. 3(d)]. To study the critical field behavior at lower temperatures, we carried out a high magnetic field measurement up to 35.5 T on 6 ML Pb film [Fig. 3(c)]. The Pauli limit for 6 ML Pb film is estimated by the equation $B_p = \Delta_0/\sqrt{2}\mu_B = 14.7$ T [13,14], since $2\Delta_0/k_B T_c$ remains a constant of about 4.4 for ultrathin crystalline Pb films [3]. The upper critical field exceeds 35.5 T at 2.8 K, which is already far beyond the Pauli limit $B_p$. Fig. 3(d) summarizes the in-plane critical magnetic field $B_c$ normalized by $B_p$ as a function of reduced temperature $T/T_c$ for 4-6 ML Pb films. Based on the Gor'kov Green's function technique [27], we develop a microscopic model for the Zeeman-protected superconductivity to determine the critical field $B_c(T)$ in terms of effective Zeeman-type SOI $\widetilde{\beta_{SO}}$:

$$\ln\left(\frac{T}{T_c}\right) + \frac{\mu_B^2 B^2}{\widetilde{\beta_{SO}}^2 + \mu_B^2 B^2} Re\left[\psi\left(\frac{1}{2} + \frac{i\sqrt{\widetilde{\beta_{SO}}^2 + \mu_B^2 B^2}}{2\pi k_B T}\right) - \psi\left(\frac{1}{2}\right)\right] = 0 \quad (1)$$

where $\psi(x)$ is the digamma function, $\widetilde{\beta_{SO}} = \beta_{SO}/(1 + \frac{\hbar}{2\pi k_B T_c \tau_0})$ is the effective Zeeman-type SOI considering spin-independent scattering and $\tau_0$ is mean free time (more details are shown in the Supplementary Information [28]).

The experimental data of $B_c/B_p$ versus $T/T_c$ can be quantitatively fitted by equation (1). The fitting procedure gives effective Zeeman-type SOI $\widetilde{\beta_{SO}} =$ 3.54 meV and 4.24 meV for 6 ML and 4 ML Pb films. The in-plane critical field $B_c$ increases with reducing film thickness, which yields larger $\widetilde{\beta_{SO}}$ for thinner films. Since the in-plane inversion symmetry is preserved in both free-standing Pb atomic layers and Si substrate, the Zeeman-type SOI and large parallel critical field observed in our system should originate from the interface between them.

The SIC phase is a special surface reconstruction of Pb on Si (111) substrate with a Pb coverage of 4/3 ML [5]. To be specific, each unit cell of SIC phase consists of four Pb atoms (the purple circles) located on three surface Si (the blue circles) atoms [Fig. 3(e)]. Three Pb atoms (marked by 2) form a trimer, which has covalent bonds with the underlying Si substrate. The other one (marked by 1) locates at the center of the trimer without bonding to the Si atoms. Metallic bonds are formed within the Pb layers. Distinct from the hexagonal close packed structure in free-standing Pb atomic layer, the Pb layer in SIC phase is apparently distorted since the distance between Pb1 and Pb3 $D_{13}$ (from the nearest unit cell) is about 20% larger than that between Pb1 and Pb2 $D_{12}$ [Fig. 3(e)] [29]. The special structure of SIC on Si(111) substrate can give rise to Zeeman-type SOI, which has been demonstrated by DFT calculation [30] and spin-resolved angular-resolved photoemission spectroscopy (ARPES) measurement [31].

The lattice distortion in SIC phase can extend to the neighboring Pb layers. Our *ab initio* calculations show that the similar lattice distortion, evaluated by $D_{13}/D_{12}$, still exists in the neighboring Pb layers grown on SIC phase ($D_{13}$ is 3.0% larger than $D_{12}$ for 1st Pb layer, see Supplementary table S1 [28] for details). Lattice distortion breaks the in-plane inversion symmetry and contributes to the local in-plane crystal field, inducing Zeeman-type SOI in ultrathin lead film on SIC. The spin splitting in Pb film on SIC phase has been further confirmed by band structure calculation. We calculate the band structure of 4 ML Pb film on SIC and 5 ML free-standing Pb film for comparison (Fig. S4), where 5 ML Pb film can be regarded as 4 ML Pb film/SIC system without lattice distortion. Pronounced spin splitting of the energy bands across the Fermi surface is obtained in 4 ML Pb/SIC system, especially around the K/-K point, which is much larger than that of 5 ML free-standing Pb film as expected. Therefore, the spin splitting in 4 ML Pb/SIC system can be mainly attributed to the lattice distortion and is very likely to be Zeeman-type. Due to the time reversal symmetry, the spin polarization of electrons with momenta at k and –k has opposite direction. Formation of Cooper pairs between electrons with opposite momentum and spin polarizations is thus favored in this situation. Zeeman-type SOI locks the spins of Cooper pairs in the out-of-plane orientation and prevents the alignment of the spin to the external in-plane magnetic field when $\mu_B B < \widetilde{\beta_{SO}}$, and the superconductivity is protected under large parallel magnetic field far beyond the Pauli limit, i.e. Zeeman-protected superconductivity (Fig. S5).

To further confirm and investigate Zeeman-protected superconductivity in ultrathin crystalline lead film, we measured the $R_{sq}(B)$ curves of a 6 ML sample in pulsed parallel magnetic field up to 47 T [Fig. 4(a)]. The superconductivity survives under a high parallel magnetic field of 40 T at 1.7 K. In Fig. 4(b), the temperature dependence of in-plane critical magnetic field, defined as the field corresponding to 50% $R_n$, is fitted by equation (1) with effective Zeeman-type SOI $\widetilde{\beta_{SO}} = 3.01$meV (close to the value measured in steady high field for another 6 ML sample, as shown in Fig. 3(d)), suggesting Zeeman-protected superconductivity is the underlying reason for the enhancement of $B_c$ in 6 ML Pb film.

We then discuss other possible origins that may contribute to a high parallel critical field in 2D superconductors, such as spin-triplet pairing, spin-orbit impurity scattering and Rashba-type SOI. In ultrathin crystalline Pb films, the out-of-plane critical field $B_{c2}(T)$ can be well fitted by WHH theory, indicating conventional *s*-wave pairing [Fig. S2]. Thus, the contribution from potential spin-triplet pairing due to SOI can be neglected. The microscopic Klemm-Luther-Beasley (KLB) theory [15] points out that the effect of spin paramagnetism can be reduced by spin-orbit impurity scattering and hence the parallel critical field can be enhanced beyond the Pauli limit. The KLB formula, however, cannot describe the observed relationship between $B_c/B_p$ and $T/T_c$ (the inset of Fig. 4(c)), indicating spin-orbit impurity scattering alone cannot account for the observed large $B_c$. We develop a microscopic theory considering both Zeeman-type SOI and spin-orbit scattering (the microscopic derivation is given in the Supplementary Information [28]):

$$\ln\left(\frac{T}{T_c}\right) + 2 \cdot Re\left\{\frac{(\mu_B B)^2}{(x_2-x_1)(x_2-x_3)}\left[\psi\left(\frac{1}{2}-\frac{x_1}{2\pi k_B T}\right) - \psi\left(\frac{1}{2}-\frac{x_2}{2\pi k_B T}\right)\right]\right\} = 0 \quad (2)$$

where $\psi(x)$ is the digamma function, and $x_1, x_2, x_3$ is the solution of the following equations:

$$x^3 + \frac{2}{3}\frac{\hbar}{\tau_{SO}}x^2 + \left(\widetilde{\beta_{SO}}^2 + \mu_B^2 B^2\right)x + \frac{2}{3}\frac{\hbar \widetilde{\beta_{SO}}^2}{\tau_{SO}} = 0$$

$$Im(x_1) = 0$$

with effective Zeeman-type SOI $\widetilde{\beta_{SO}} = \beta_{SO}/(1+\frac{\hbar}{2\pi k_B T_c \tau_0})$ and mean free time $\tau_{SO}$ for spin-orbit impurity scattering and mean free time $\tau_0$ for spin-independent impurity scattering. The fitting curve gives $\hbar/\tau_{SO} \approx 0$ meV, which is very small compared to the effective Zeeman-type SOI $\widetilde{\beta_{SO}} = 3.01$ meV. Therefore, the Zeeman-type SOI is the dominating mechanism of the enhanced $B_c$.

Figure 4(d) presents the fitting results of the in-plane critical field $B_c(T)$ data for 6 ML Pb film by a microscopic model with both the Zeeman-type and Rashba-type SOI (see more details and microscopic derivation in the Supplementary Information [28]), in which $B_c$ as a function of $\frac{T}{T_c}$ is determined by both the effective Rashba-type SOI $\widetilde{\alpha_R k_F}$ and Zeeman-type SOI $\widetilde{\beta_{SO}}$:

$$\ln\left(\frac{T}{T_c}\right) + \frac{1}{2}\left[1 - \frac{2(\widetilde{\alpha_R k_F})^2 + \widetilde{\beta_{SO}}^2 - \mu_B^2 B^2}{\rho_+^2 - \rho_-^2}\right] Re\left[\psi\left(\frac{1}{2} + \frac{i\rho_+}{2\pi k_B T}\right) - \psi\left(\frac{1}{2}\right)\right] + \frac{1}{2}\left[1 + \frac{2(\widetilde{\alpha_R k_F})^2 + \widetilde{\beta_{SO}}^2 - \mu_B^2 B^2}{\rho_+^2 - \rho_-^2}\right] Re\left[\psi\left(\frac{1}{2} + \frac{i\rho_-}{2\pi k_B T}\right) - \psi\left(\frac{1}{2}\right)\right] = 0 \quad (3)$$

Where $\psi(x)$ is the digamma function, and the arguments are defined as

$$2\rho_\pm \equiv \sqrt{(\mu_B B + \widetilde{\alpha_R k_F})^2 + (\widetilde{\alpha_R k_F})^2 + \widetilde{\beta_{SO}}^2} \pm \sqrt{(\mu_B B - \widetilde{\alpha_R k_F})^2 + (\widetilde{\alpha_R k_F})^2 + \widetilde{\beta_{SO}}^2}$$

with effective Zeeman-type SOI $\widetilde{\beta_{SO}} = \beta_{SO}/(1 + \frac{\hbar}{2\pi k_B T_c \tau_0})$ and effective Rashba-type SOI $\widetilde{\alpha_R k_F} = \alpha_R k_F/\sqrt{2}(1 + \frac{\hbar}{2\pi k_B T_c \tau_0})$. The result of equation (3) is consistent with the previous theory obtained under clean limit [22].

As shown in Fig. 4(d), the theoretical curve is in good agreement with experimentally observed in-plane critical field (normalized $B_c/B_p$ versus $T/T_c$) with $\widetilde{\beta_{SO}} = 3.16$ meV and $\widetilde{\alpha_R k_F} = 0.22$ meV for 6 ML Pb film. In Zeeman-protected superconducting systems, the spin orientation is polarized to the out-of-plane direction by Zeeman-type SOI, while Rashba-type SOI weakens the Zeeman-protection mechanism by tilting the spin of the electron to the in-plane direction, which can be more easily affected by the parallel magnetic field. To clarify the influence of Rashba-type SOI, we display a set of theoretical curves with a fixed effective Zeeman-type SOI $\widetilde{\beta_{SO}}$ of 3.16 meV and increasing effective Rashba-type SOI $\widetilde{\alpha_R k_F}$ from 0.22 meV to 1.26 meV (Fig. 4(d)). The Rashba-type SOI bends down the curves at low temperatures and gives rise to a relatively small in-plane critical field, which qualitatively deviates from the quasi-linear temperature dependence of our measured data. Besides, a special case with only large effective Rashba-type SOI is also considered (the olive line, $\widetilde{\alpha_R k_F} = 30$ meV, $\widetilde{\beta_{SO}} = 0$ meV), where $B_c$ is far below the experimental data, indicating Rashba-type SOI alone cannot account for the observed large in-plane critical field.

As a conclusion, we systematically investigate the transport properties of ultrathin crystalline Pb films at low temperatures and high magnetic fields. Under parallel magnetic field, superconductivity protected by intrinsic Zeeman-type SOI survives at high fields far beyond the Pauli limit. We demonstrate that the interface between Pb film and Si substrate plays a key role in breaking in-plane inversion symmetry and hence inducing Zeeman-protected superconductivity in Pb films. Our work reveals that in epitaxial heterostructures the interface modulated SOI has profound influence on the superconducting pairing, and may provide a promising platform to study unconventional superconductivity [32].

We thank Ying Xing, Yanan Li, Jinglei Zhang, Youmin Zou, Jiezun Ke, Liang Li, Cheng Chen, Yue Tang for the help in transport measurements. This work was financially supported by the National Basic Research Program of China (Grant No. 2013CB934600, No. 2017YFA0303302, No. 2016YFA0301004, No. 2015CB921102, No. 2017YFA0303301 and No.2017YFA0304600), the National Natural Science Foundation of China (Grant No.11774008, No. 11534001 and No. 11674028), and the Open Project Program of the Pulsed High Magnetic Field Facility (Grant No. PHMFF2015002) at the Huazhong University of Science and Technology, and the Key Research Program of the Chinese Academy of Sciences (Grant No. XDPB08-1). H. -W. Liu also acknowledges support from the Fundamental Research Funds for the Central Universities.

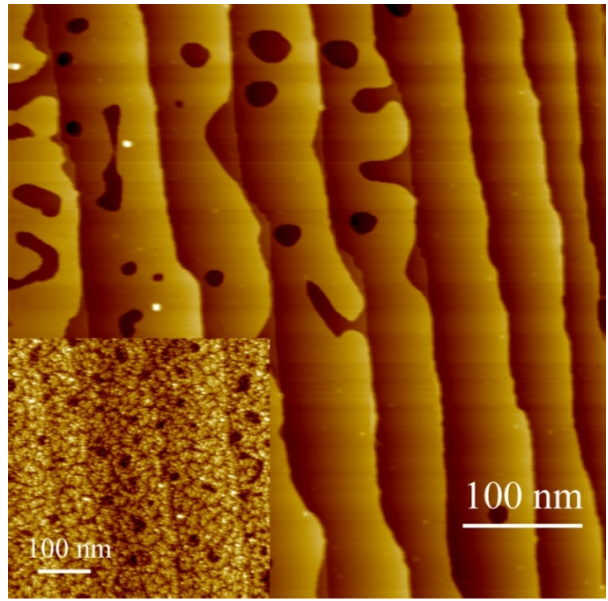

FIG. 1. A typical STM image of 4 ML Pb film. (500 nm × 500 nm, sample bias 2.0 V, tunneling current 500 pA) Inset: STM image of Si capping layer on 4 ML Pb film. (500 nm × 500 nm, sample bias 3.0 V, tunneling current 500 pA)

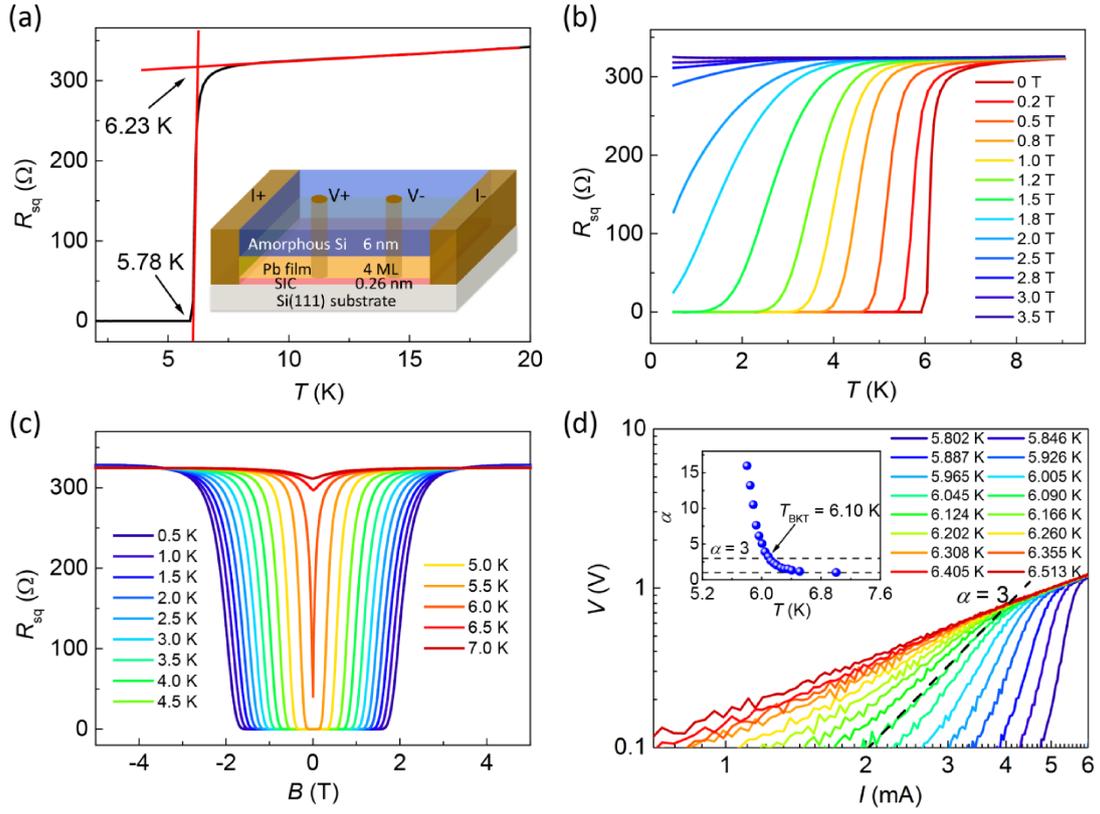

FIG. 2. Transport properties of 4 ML Pb film (Sample 1). (a) Temperature dependence of sheet resistance $R_{sq}$ at zero magnetic field, showing $T_c^{onset}$ = 6.29 K and $T_c^{zero}$ = 5.73 K. The inset is a schematic for standard four-electrode transport measurements. (b) $R_{sq}(T)$ measured under various perpendicular magnetic fields up to 3.5 T. (c) The field dependence of $R_{sq}$ at different temperatures ranging from 0.5 to 7.0 K. (d) $V(I)$ characteristics at various temperatures from 5.802 to 6.513 K at B = 0 T plotted on a double-logarithmic scale. The black line corresponds to $V \propto I^3$ dependence. Inset: power-law exponent α ($V \propto I^\alpha$) as a function of temperature, showing $T_{BKT}$ = 6.10 K.

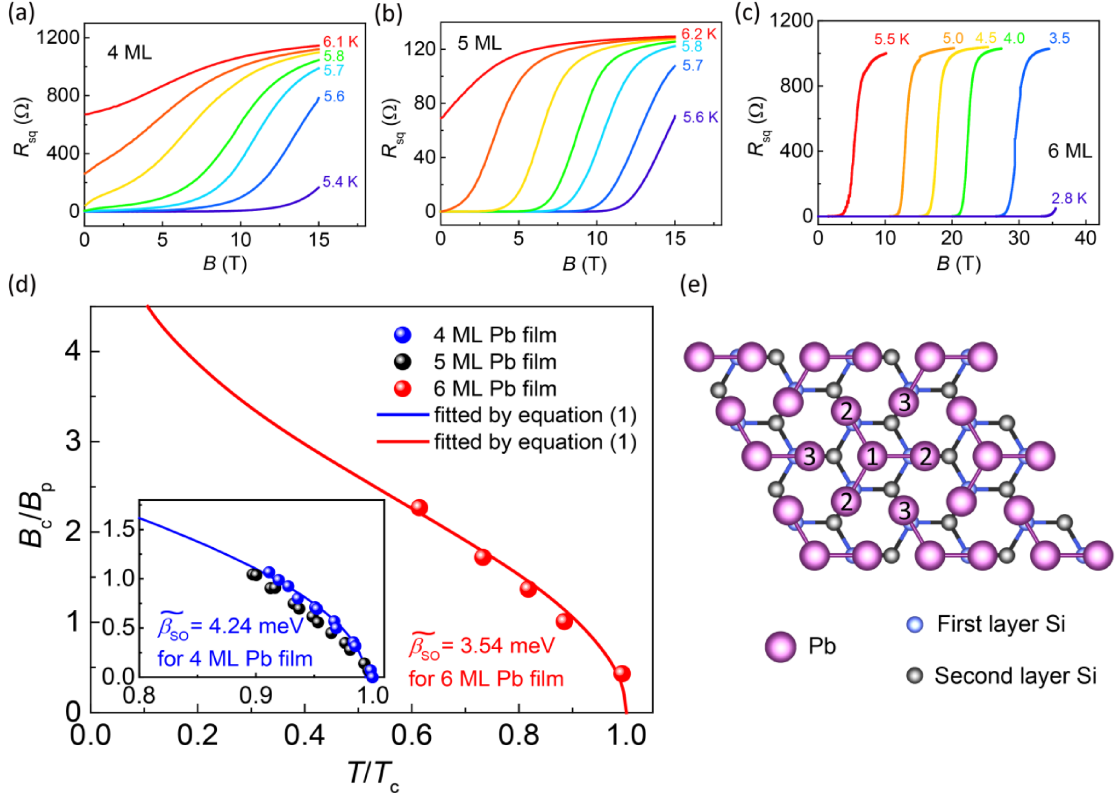

FIG. 3. Experimental evidence for Zeeman-protected superconductivity in ultrathin (4-6 ML) Pb films. (a)-(c) Parallel magnetic field dependence of $R_{sq}$ at various temperatures for (a) 4 ML (Sample 1), (b) 5 ML (Sample 2) and (c) 6 ML (Sample 3) Pb films. At 2.8 K or lower temperatures, superconductivity survives under high magnetic fields up to 35.5 T for 6 ML Pb film. (d) $B_c$ normalized by Pauli limit $B_p$ as a function of reduced temperature $T/T_c$ for 6 ML Pb film. Here, $B_c$ is determined as the magnetic field required to reach 50% of normal state sheet resistance $R_n$. The measured $B_c$ can be well fitted by the theoretical curve for Zeeman-protected superconductivity (equation (1)), giving effective Zeeman-type SOI $\widetilde{\beta_{SO}} = 3.54$ meV in 6 ML Pb film. Inset: Normalized critical field $B_c/B_p$ for 4 ML, 5 ML Pb films measured in PPMS with magnetic fields up to 15 T. The blue line is a fitting curve using equation (1), giving $\widetilde{\beta_{SO}} = 4.24$ meV in 4 ML Pb film. (e) The top view of the crystal structure of the SIC phase. Each unit cell consists of four Pb atoms, with three Pb atoms (marked as 2) forming a trimer and the other one (marked as 1) located at the center of the trimer. The distance between Pb1 and Pb3 (from the nearest unit cell) is much larger than that between Pb1 and Pb2, and the in-plane inversion symmetry is broken in this system.

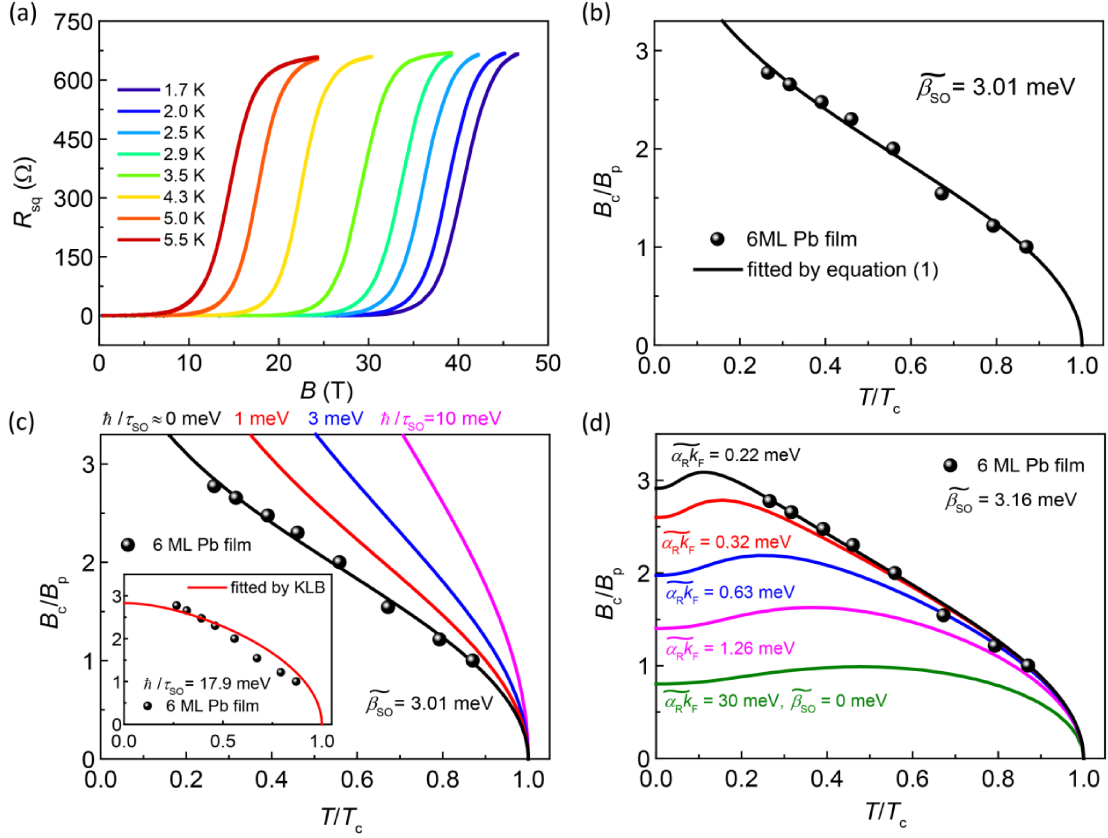

FIG. 4. Pulsed high magnetic field measurement for another 6 ML Pb film (Sample 4). (a) $R_{sq}(B)$ curves at various temperatures measured at parallel magnetic fields up to 47 T. The superconductivity survives under a high parallel magnetic field of 40 T at 1.7 K. (b) The in-plane critical field $B_c/B_p$ as a function of normalized temperature $T/T_c$ for 6 ML Pb film shows a good agreement with the microscopic theory of Zeeman-protected superconductivity (the black solid line, equation (1)), yielding an effective Zeeman-type SOI $\widetilde{\beta_{SO}} = 3.01$ meV. (c) The $B_c/B_p$ ($T/T_c$) curve is fitted based on a microscopic theory considering both effective Zeeman-type SOI and spin-orbit scattering (the black solid line, equation (2)) and gives $\hbar/\tau_{SO} \approx 0$ meV. The red, blue and magenta lines indicate the theoretical curves with a fixed $\widetilde{\beta_{SO}} = 3.01$ meV and increasing $\hbar/\tau_{SO}$ from 1 to 10 meV. The inset: The relationship between $B_c/B_p$ and $T/T_c$ cannot be fitted by the KLB formula, which describes the in-plane critical field enhanced only by spin-orbit impurity scattering. (d) Theoretical fitting of $B_c/B_p$ as a function of $T/T_c$ using a fixed effective Zeeman-type SOI and increasing effective Rashba-type SOI. The black curve presents the best fit of the experimental data with effective Zeeman-type SOI $\widetilde{\beta_{SO}} = 3.16\ \text{meV}$ and effective Rashba-type SOI $\widetilde{\alpha_R k_F} = 0.22\ \text{meV}$. The olive line indicates a special case with only effective Rashba-type SOI ($\widetilde{\alpha_R k_F} = 30\ \text{meV}, \widetilde{\beta_{SO}} = 0$), which is far below the experimental results.

# Supplementary Information

# Interface induced Zeeman-protected superconductivity in ultrathin crystalline lead films

Yi Liu[1,2,†], Ziqiao Wang[1,2,†], Xuefeng Zhang[1,2], Chaofei Liu[1,2], Yongjie Liu[3], Zhimou Zhou[1,2], Junfeng Wang[3], Qingyan Wang[1,2,4], Yanzhao Liu[1,2], Chuanying Xi[5], Mingliang Tian[5], Haiwen Liu[6,*], Ji Feng[1,2], X. C. Xie[1,2] and Jian Wang[1,2,7,*]

[1]*International Center for Quantum Materials, School of Physics, Peking University, Beijing 100871, China.*
[2]*Collaborative Innovation Center of Quantum Matter, Beijing 100871, China.*
[3]*Wuhan National High Magnetic Field Center, Huazhong University of Science and Technology, Wuhan 430074, China.*
[4]*Institute of Physics, Chinese Academy of Sciences, Beijing 100190, China.*
[5]*High Magnetic Field Laboratory, Chinese Academy of Sciences, Hefei 230031, Anhui, China.*
[6]*Center for Advanced Quantum Studies, Department of Physics, Beijing Normal University, Beijing 100875, China.*
[7]*CAS Center for Excellence in Topological Quantum Computation, University of Chinese Academy of Sciences, Beijing 100190, China*

†These authors contributed equally to this work.
*Corresponding authors: Haiwen Liu (haiwen.liu@bnu.edu.cn) and Jian Wang (jianwangphysics@pku.edu.cn)

**Contents**

I. Sample growth and transport measurement
II. The first-principles calculations of crystal and electronic structure

III. The microscopic model of Zeeman-type spin-orbit interaction protected superconductivity

IV. Comparison between Zeeman-type SOI mechanism and GL formula

V. Figures and Tables

## I. Sample growth and transport measurement

The ultrathin crystalline Pb films were grown in ultrahigh vacuum MBE chamber (Omicron) with base pressure better than $1\times10^{-10}$ mbar. The Si(111) – 7×7 reconstruction was prepared by cycles of flashing Si(111) substrate at $T \sim 1400$ K. The SIC phase was acquired by deposition of ~ 1.5 ML Pb from a Knudsen cell at room temperature and subsequent annealing at $T \sim 573$ K for 30 sec. The ultrathin Pb films were prepared by depositing pure Pb atoms on SIC phase or amorphous wetting layer with a rate of ~ 0.2 ML/min at $T_{substrate} \sim 150$ K. Film growth was monitored by reflection high-energy electron diffraction (RHEED) and characterized by STM.

For ex situ transport measurements, Pb films were protected and immobilized by amorphous Si capping layer with thickness of 6 nm. The resistance and magnetoresistance were measured with the standard four-probe method in a commercial Physical Property Measurement System (Quantum Design, PPMS-16) with the Helium-3 option for temperature down to 0.5 K under perpendicular magnetic field. The films were mounted on the rotation holder for measurement in parallel magnetic field with an uncertainty within 0.1°. The contact between electrodes and films are made by pressing indium on the sample surface. The current electrodes are narrow strips along the width of the sample to assure the current is uniform in the films. The steady high magnetic field measurements were carried out using standard ac lock-in technique (13.7 Hz) with a He-3 fridge and water-cooling magnet up to 35.5 T in the National High Magnetic Field Lab of China at Hefei. The pulsed high field measurements were performed under a parallel magnetic field up to 47 T with a base temperature of 1.7 K in National High Magnetic Field Center at Wuhan. Standard lock-in technique was used during the measurements with a current excitation of 500 μA at 100 kHz.

## II. The first-principles calculations of crystal and electronic structure

Density functional theory calculations were performed using the Vienna ab initio simulation package (VASP) [1], with projector augmented wave (PAW) pseudopotentials [2,3]. The exchange-correlation functional in the Perdew-Burke-Ernzerhof (PBE) form of generalized-gradient approximation (GGA) [4] is used to investigate the geometry and electronic structure of 4 ML Pb films grown on SIC on Si(111) substrate. The Kohn-Sham single-particle wavefunctions were expanded in the plane wave basis set with a kinetic energy truncation at 400 eV. The Brillouin zone sampling was performed on a $9 \times 9 \times 1$ k-grid with Gaussian type broadening of 0.05 eV. The crystal structure is relaxed with a conjugate gradient method until Hellmann-Feynman forces on each atom are less than 0.01 eV/Å. Spin-orbit interaction is taken into account in the band structure calculation.

To perform the first-principles calculations, we put 4 ML Pb(111) film on SIC phase on Si(111) substrate. The Pb/SIC/Si(111) system is modeled by a periodic slab which consists of six layers of Si passivated by H at the bottom, and SIC phase, Pb film on the top of the Si(111) substrate. The bottom double layers of Si atoms and H atoms are kept fixed to simulate the substrate environment. We put 4 ML 2×2 Pb(111) films on SIC/Si(111) substrate using the lattice constant of SIC phase, which has an acceptable mismatch less than 5% compared to the freestanding Pb film. We employed a vacuum region of larger than 20 Å to ensure sufficient separation between the periodic images of Pb/SIC/Si(111) system.

## III. The microscopic model of Zeeman-type spin-orbit interaction protected superconductivity

Based on the low mobility of our Pb thin films in the normal state, the system belongs to the family of dirty superconductivity. We apply the Gor'kov–Abrikovsov theory [5], and give the Gor'kov Green's function in real space as follows [5,6]. The Gap function reads:

$$\Delta_\omega(\vec{r} - \vec{r}') = \delta^3(\vec{r} - \vec{r}')\Delta + \iint d^3\vec{r} d^3\vec{r}' \langle V(\vec{r} - \vec{r_1}) F_\omega(\vec{r_1} - \vec{r_2}) V(\vec{r_2} - \vec{r}') \rangle, \text{(S1)}$$

with $V(\vec{r})$ denoting the non-magnetic impurity scattering and $\Delta$ denoting the superconductor gap. The anomalous Green's function $F_\omega(\vec{r} - \vec{r}')$ is defined by:

$$F_\omega(\vec{r} - \vec{r}') = \iint d^3\vec{r_1} d^3\vec{r_2} G^n_{\omega,\sigma}(\vec{r} - \vec{r_1}) \Delta_\omega(\vec{r_1} - \vec{r_2}) G^n_{-\omega,-\sigma}(\vec{r}' - \vec{r_2}), \text{(S2)}$$

And the transition temperature $T$ obeys [14]:

$$\ln\left(\frac{T_c}{T}\right) = k_B T \sum_n \left[\frac{\pi}{|\omega|} - \frac{1}{2} tr\left\{\frac{F_\omega(0)}{N(0)\Delta}\right\}\right], \text{(S3)}$$

and the Matsubara frequency is $|\omega| = (2n + 1)\pi k_B T$ with Boltzmann constant $k_B$ and $N(0)$ is the density of states. We consider the in-plane magnetic field has no orbital effect, thus the system has translation symmetry in the Born approximation. After transforming the problem into momentum space, the anomalous Green's function can be represented as:

$$F_\omega(0) = \int F_\omega(\vec{p}) d^3\vec{p} \equiv \frac{\Delta}{k_B T} \int d^2\hat{p} S_\omega(\hat{p}) = \frac{\Delta N(0)}{k_B T} S_\omega(\hat{p}), \text{(S4)}$$

Here we assume $S_\omega(\hat{p})$ to be isotropic, which can be justified in the following part. We introduce an integral kernel function as:

$$S_\omega(\hat{p}) \equiv \frac{k_B T}{\Delta} \int d\xi F_\omega(\vec{p}), \text{(S5)}$$

with $\xi = \frac{p^2}{2m} - E_F$. Thus, the relation for transition temperature $T$ in Eq. (S3) reduces to:

$$\ln\left(\frac{T_c}{T}\right) = k_B T \sum_n \left[\frac{\pi}{|\omega|} - \frac{1}{2k_B T} tr S_\omega\right]. \text{(S6)}$$

We firstly neglect the disorder influence in Eq. (S1), and the bare anomalous Green's function reads:

$$F^0_\omega(\vec{r} - \vec{r}') = \Delta \iint d^3\vec{r_1} G^n_{\omega,\sigma}(\vec{r} - \vec{r_1}) G^n_{-\omega,-\sigma}(\vec{r}' - \vec{r_1}). \text{(S7)}$$

We consider the simplest model with Zeeman-type spin splitting in *z*-direction and in-plane magnetic field deduced spin splitting in x-direction:

$$H_{\vec{p}} = \frac{p^2}{2m} - \beta_{SO}\sigma_z\tau_z - \mu_B B\sigma_x, \quad (S8)$$

here $\sigma$ and $\tau$ denotes the real spin and valley index respectively.

*Section 3.1 Zeeman-type SOI protected superconductivity with spin-independent scattering.*

The spin-independent scattering disorder gives a finite life time $\frac{\hbar}{\tau_0} = 2\pi n_i N(0)V^2$ with $n_i$ denoting the disorder concentration. And the diagonal Green's function under influence of non-magnetic disorder scattering reads:

$$G^n_{\omega,\sigma}(\vec{p}) = \frac{1}{i|\omega| + \frac{i\hbar}{2\tau_0} + (\xi - \beta_{SO}\sigma_z\tau_z - \mu_B B\sigma_x I_2)\cdot sgn(\omega)}. \quad (S9)$$

Next, we calculate the bare integral kernel $S^0_\omega$. Firstly, we introduce a useful integral identity for further use:

$$\int_{-\infty}^{\infty} \frac{dx}{(x - i + A\sigma_z\tau_z + B\sigma_x I_2)(x + i + A\sigma_z\tau_z - B\sigma_x I_2)} = \frac{\pi(A^2 + 1)}{(A^2 + 1) + i(B\sigma_x I_2 - AB\sigma_y\tau_z)}. \quad (S10)$$

Thus the bare integral kernel $S^0_\omega$ corresponding to $F^0_\omega$ in Eq. (S8) reads:

$$S^0_\omega(\hat{p}) = \frac{\pi k_B T}{|\omega| + \frac{\hbar}{2\tau_0}} \frac{\beta^2_{SO} + \left(|\omega| + \frac{\hbar}{2\tau_0}\right)^2}{\beta^2_{SO} + \left(|\omega| + \frac{\hbar}{2\tau_0}\right)^2 - i\left[\beta_{SO}\mu_B B\sigma_y\tau_z + \mu_B B\left(|\omega| + \frac{\hbar}{2\tau_0}\right)\sigma_x I_2\right]\cdot sgn(\omega)}, \quad (S11)$$

Considering the disorder influence in Eq. (S2), the full integral kernel $S_\omega$ satisfies:

$$S_\omega(\hat{p}) = S^0_\omega(\hat{p})\left[1 + \frac{n_i N(0)V^2}{k_B T}S_\omega(\hat{p})\right]. \quad (S12)$$

Then the full kernel function $S_\omega$ is obtained:

$$S_\omega(\hat{p}) = \frac{\pi k_B T}{|\omega|} \frac{\beta^2_{SO} + \left(|\omega| + \frac{\hbar}{2\tau_0}\right)^2}{\beta^2_{SO} + \left(|\omega| + \frac{\hbar}{2\tau_0}\right)^2 - i\left(1 + \frac{\hbar}{2|\omega|\tau_0}\right)\left[\beta_{SO}\mu_B B\sigma_y\tau_z + \mu_B B\left(|\omega| + \frac{\hbar}{2\tau_0}\right)\sigma_x I_2\right]\cdot sgn(\omega)}, \quad (S13)$$

Then we can determine the relation between the upper critical field $B_c(T)$ and temperature $T$:

$$\ln\left(\frac{T_c}{T}\right) = k_B T \sum_{n=-\infty}^{+\infty} \left[\frac{\pi}{|\omega|} - \frac{1}{2k_B T} tr S_\omega(\hat{p})\right]$$

$$= \pi k_B T \sum_n \frac{1}{|\omega|} \frac{\mu_B^2 B^2}{\frac{\beta^2_{SO}}{\left(1 + \frac{\hbar}{2|\omega|\tau_0}\right)^2} + \omega^2 + \mu_B^2 B^2}. \quad (S14)$$

We firstly check several special cases for Eq. (S14):

(i), setting the in-plane magnetic field $B = 0$, one obtains $T = T_c$, indicating the Zeeman-type SOI cannot solely influence $T_c$.

(ii), setting the Zeeman-type SOI $\beta_{SO} = 0$, one obtains the upper critical field $B_c(T)$ identical to previous results with $\psi(x)$ denoting the digamma function [6]:

$$ln\left(\frac{T_c}{T}\right) = \pi k_B T \sum_n \frac{1}{|\omega|} \frac{\mu_B^2 B^2}{\omega^2 + \mu_B^2 B^2}$$

$$= \pi k_B T \sum_n \frac{1}{|\omega|} \frac{\mu_B^2 B^2}{\omega^2 + \mu_B^2 B^2} = Re\left[\psi\left(\frac{1}{2} + \frac{i\mu_B B}{2\pi k_B T}\right) - \psi\left(\frac{1}{2}\right)\right]. \text{ (S15)}$$

(iii), considering the clean limit by setting $\frac{\hbar}{2|\omega|\tau_0} = 0$, the upper critical field $B_c(T)$ satisfies:

$$ln\left(\frac{T_c}{T}\right) = \pi k_B T \sum_n \frac{1}{|\omega|} \frac{\mu_B^2 B^2}{\beta_{SO}^2 + \mu_B^2 B^2 + \omega^2}$$

$$= \frac{\mu_B^2 B^2}{\beta_{SO}^2 + \mu_B^2 B^2} Re\left[\psi\left(\frac{1}{2} + \frac{i\sqrt{\beta_{SO}^2 + \mu_B^2 B^2}}{2\pi k_B T}\right) - \psi\left(\frac{1}{2}\right)\right]. \text{ (S16)}$$

Although $B_c(T)$ is not influenced by disorder scattering for case (i) and case (ii), $B_c(T)$ can surely be altered by disorder scattering with coexistence of both Zeeman-type SOI and in-plane magnetic field. We assume the approximation in Eq. (S14) that $1 + \frac{\hbar}{2|\omega|\tau_0} \approx 1 + \frac{\hbar}{2\pi k_B T_c \tau_0}$, and introduce new parameter $\widetilde{\beta_{SO}} = \frac{\beta_{SO}}{1 + \frac{\hbar}{2\pi k_B T_c \tau_0}}$. Thus, the upper critical field $B_c(T)$ for the Zeeman-type SOI protected superconductivity in the dirty limit reads:

$$ln\left(\frac{T_c}{T}\right) = \pi k_B T \sum_n \frac{1}{|\omega|} \frac{\mu_B^2 B^2}{\widetilde{\beta_{SO}}^2 + \omega^2 + \mu_B^2 B^2}$$

$$= \frac{\mu_B^2 B^2}{\widetilde{\beta_{SO}}^2 + \mu_B^2 B^2} Re\left[\psi\left(\frac{1}{2} + \frac{i\sqrt{\widetilde{\beta_{SO}}^2 + \mu_B^2 B^2}}{2\pi k_B T}\right) - \psi\left(\frac{1}{2}\right)\right], \text{ (S17)}$$

and here $\widetilde{\beta_{SO}} = \frac{\beta_{SO}}{1 + \frac{\hbar}{2\pi k_B T_c \tau_0}}$. Equation (S17) is the central result of this section. The disorder scattering can give rise to band broadening effect, reduce the Zeeman-type SOI spin splitting, and ultimately weaken the Zeeman-type SOI protected superconductivity to some extent as shown in equation (S17). Comparing the fitting result $\widetilde{\beta_{SO}}$ shown in Figure 3 of the main text with the band calculation of spin splitting in Fig. S3 for 4 ML Pb film, we estimate $\frac{\hbar}{2\pi k_B T_c \tau_0}$ is of order one, and the superconductor belongs to the dirty limit.

*Section 3.2 Zeeman-type SOI protected superconductivity with spin-flipping scattering*

In the following part, we consider both the spin-independent scattering and the spin-flipping scattering. The characteristic time satisfies $\tau^{-1} = \tau_0^{-1} + \tau_{so}^{-1}$, and $\tau, \tau_0, \tau_{so}$ represents the mean free time for total scattering, spin-independent scattering and spin-flipping scattering, respectively. Considering the spin-flipping scattering, we obtain the full integral kernel $S_\omega$ obeys the following form:

$$S_\omega(\hat{p}) \equiv S_\omega^{(1)}(\hat{p}) + \left[S_\omega^{(2)}(\hat{p})\sigma_x I_2 + S_\omega^{(3)}(\hat{p})\sigma_y \tau_z\right]\vec{\sigma} \cdot \hat{H}\, sgn(\omega). \text{ (S18)}$$

Considering the spin-flipping scattering, $S_\omega$ satisfies the equation:

$$S_\omega(\hat{p}) = S_\omega^0(\hat{p}) \left\{1 + \frac{\tau^{-1} \cdot S_\omega(\hat{p})}{2\pi k_B T} - \frac{4}{3}\tau_2^{-1}\left[S_\omega^{(2)}(\hat{p})\sigma_x I_2 + S_\omega^{(3)}(\hat{p})\sigma_y \tau_z\right]/2\pi k_B T\right\}, \text{ (S19)}$$

with the bare integral kernel $S_\omega^0(\hat{p})$ given in Eq. (S11), and then the full integral kernel $S_\omega(\hat{p})$ simplifies into the following form:

$$S_\omega(\hat{p}) = \frac{\pi k_B T}{|\omega|} \cdot \frac{\widetilde{\beta_{SO}^2} + |\omega|^2 + i\mu_B B\left[\widetilde{\beta_{SO}}\sigma_y\tau_z + |\omega|\sigma_x I_2\right] \cdot sgn(\omega)}{\widetilde{\beta_{SO}^2} + |\omega|^2 + (\mu_B B)^2}\left\{1 - \frac{4}{3}\tau_{so}^{-1}\frac{S_\omega^{(2)}(\hat{p})\sigma_x I_2 + S_\omega^{(3)}(\hat{p})\sigma_y\tau_z}{2\pi k_B T}\right\}. \text{ (S20)}$$

Here, we use the notation $\widetilde{\beta_{SO}} = \frac{\beta_{SO}}{1 + \frac{\hbar}{2\pi k_B T \tau_0}}$. The equation for $S_\omega^{(i)}(\hat{p})$ reads:

$$S_\omega^{(1)}(\hat{p}) = \frac{\pi k_B T}{|\omega|} \cdot \frac{\widetilde{\beta_{SO}^2} + |\omega|^2 - i\frac{4}{3}\tau_{so}^{-1}\mu_B B \frac{|\omega| S_\omega^{(2)}(\hat{p}) + \widetilde{\beta_{SO}} S_\omega^{(3)}(\hat{p})}{2\pi k_B T}}{\widetilde{\beta_{SO}^2} + |\omega|^2 + (\mu_B B)^2}, \text{ (S21-1)}$$

$$S_\omega^{(2)}(\hat{p}) = \frac{\pi k_B T}{|\omega|} \cdot \frac{i\mu_B B|\omega| - \left(\widetilde{\beta_{SO}^2} + |\omega|^2\right)\frac{4}{3}\tau_{so}^{-1}\frac{S_\omega^{(2)}(\hat{p})}{2\pi k_B T}}{\widetilde{\beta_{SO}^2} + |\omega|^2 + (\mu_B B)^2}, \text{ (S21-2)}$$

$$S_\omega^{(3)}(\hat{p}) = \frac{\pi k_B T}{|\omega|} \cdot \frac{i\mu_B B|\omega| - \left(\widetilde{\beta_{SO}^2} + |\omega|^2\right)\frac{4}{3}\tau_{so}^{-1}\frac{S_\omega^{(3)}(\hat{p})}{2\pi k_B T}}{\widetilde{\beta_{SO}^2} + |\omega|^2 + (\mu_B B)^2}. \text{ (S21-3)}$$

The we obtain the exact form for $S_\omega^{(1)}(\hat{p})$ as follows:

$$S_\omega^{(1)}(\hat{p}) = \frac{\pi k_B T}{|\omega|} \cdot \frac{\left[\widetilde{\beta_{SO}^2} + |\omega|^2\right] \cdot \left[|\omega| + \frac{2}{3}\tau_{so}^{-1}\right]}{|\omega|\left[\widetilde{\beta_{SO}^2} + |\omega|^2 + (\mu_B B)^2\right] + \frac{2}{3}\tau_2^{-1}\left[\widetilde{\beta_{SO}^2} + |\omega|^2\right]}. \text{ (S22)}$$

Substituting Eq. (S22) into Eq. (S6), we obtain the relation for transition temperature $T$ reads:

$$\ln\left(\frac{T_c}{T}\right) = \pi k_B T \sum_n \frac{\mu_B^2 B^2}{|\omega|\left[\widetilde{\beta_{SO}^2}+|\omega|^2+(\mu_B B)^2\right]+\frac{2}{3}\tau_{so}^{-1}\left[\widetilde{\beta_{SO}^2}+|\omega|^2\right]}. \quad (S23)$$

We can find the by setting the spin-flipping scattering rate $\tau_{so}^{-1} = 0$, Eq. (S23) returns to the formula for Zeeman-type SOI protected superconductivity in Eq. (S17). Furthermore, by setting the Zeeman-type SOI $\widetilde{\beta_{SO}} = 0$, Eq. (S23) returns to the Klemm-Luther-Beasley (KLB) theory [15]. By introducing the solution of cubic equation $x^3 + \frac{2}{3}\frac{\hbar}{\tau_{SO}}x^2 + \left(\widetilde{\beta_{SO}}^2 + \mu_B^2 B^2\right)x + \frac{2}{3}\frac{\hbar \widetilde{\beta_{SO}}^2}{\tau_{SO}} = 0$, with real solution $x_1$ and complex solutions $x_2, x_3$. Eq. (S23) can be simplified into the digamma function as shown in Eq. (2) in the main text:

$$\ln\left(\frac{T_c}{T}\right) = 2 \cdot Re\left\{\frac{(\mu_B B)^2}{(x_2-x_1)(x_2-x_3)}\left[\psi\left(\frac{1}{2} - \frac{x_1}{2\pi k_B T}\right) - \psi\left(\frac{1}{2} - \frac{x_2}{2\pi k_B T}\right)\right]\right\} = 0. \quad (S24)$$

*Section 3.3 Zeeman-type SOI protected superconductivity with Rashba spin-orbit interaction.*

We extend the simplest model in section 3.1 to include both the Zeeman-type spin splitting in *z*-direction, in-plane Rashba SOI and the in-plane magnetic field induced spin splitting in x-direction at the K point and K' point:

$$H_{\vec{p},K} = \frac{p^2}{2m} - \beta_{SO}\sigma_z\tau_z - \mu_B B\sigma_x - \alpha_R(\sigma_x k_y - \sigma_y k_x), \quad (S25\text{-}1)$$

$$H_{\vec{p},K'} = \frac{p^2}{2m} - \beta_{SO}\sigma_z\tau_z - \mu_B B\sigma_x + \alpha_R(\sigma_x k_y - \sigma_y k_x). \quad (S25\text{-}2)$$

One can see that the Rashba SOI preserves the time reversal symmetry (TRS) while the in-plane magnetic field induced spin splitting breaks the (TRS). Here we introduce the second useful integral identity for further use:

$$\int_{-\infty}^{\infty} \frac{dx}{(x - i + A\sigma_z\tau_z + B\sigma_x I_2 + C\sigma_x I_2 + D\sigma_y I_2)(x + i + A\sigma_z\tau_z - B\sigma_x I_2 + C\sigma_x I_2 + D\sigma_y I_2)}$$
$$= \frac{\pi(A^2+C^2+D^2+1)}{(A^2+C^2+D^2+1)+iB[(C^2+1)\sigma_x I_2-A\sigma_y\tau_z+CD\sigma_y I_2+D\sigma_z I_2+AC\sigma_z\tau_z]}. \quad (S26)$$

Then, one can obtain the equation of transition temperature $T$ in the similar way as shown in Eq. (S9-S14). The final equation reads:

$$\ln\left(\frac{T_c}{T}\right) = \pi k_B T \sum_n \overline{\left[\frac{1}{|\omega|} \cdot \frac{2\left(\mu_B B\widetilde{\alpha_R k_F}\right)^2 \sin^2\theta + |\omega|^2(\mu_B B)^2}{|\omega|^2\left[\widetilde{\beta_{SO}}^2+2\left(\widetilde{\alpha_R k_F}\right)^2+(\mu_B B)^2+|\omega|^2\right]+2\left(\mu_B B\widetilde{\alpha_R k_F}\right)^2 \sin^2\theta}\right]}. \quad (S27\text{-}1)$$

Here, $\overline{f(\theta)} \equiv \frac{1}{2\pi}\int_0^{2\pi} f(\theta)\,d\theta$ and we only consider spin-independent scattering, and we define the effective Rashba SOI $\widetilde{\alpha_R k_F} \equiv \frac{\alpha_R k_F}{\sqrt{2}\left(1+\frac{\hbar}{2\pi k_B T \tau_0}\right)}$. for convenience. Moreover, we assume the angular average can be calculated separately, then we obtain:

$$ln\left(\frac{T_c}{T}\right) = \pi k_B T \sum_n \left[\frac{1}{|\omega|} \cdot \frac{(\mu_B B \widetilde{\alpha_R k_F})^2 + |\omega|^2 (\mu_B B)^2}{|\omega|^2 \left[\widetilde{\beta_{SO}}^2 + 2(\widetilde{\alpha_R k_F})^2 + (\mu_B B)^2 + |\omega|^2\right] + (\mu_B B \widetilde{\alpha_R k_F})^2}\right]. \quad (S27\text{-}2)$$

The Eq. (S27-2) can be simplified using digamma functions after introducing auxiliary quantities:

$$2\rho_{\pm} \equiv \sqrt{(\mu_B B + \widetilde{\alpha_R k_F})^2 + (\widetilde{\alpha_R k_F})^2 + \widetilde{\beta_{SO}}^2} \pm \sqrt{(\mu_B B - \widetilde{\alpha_R k_F})^2 + (\widetilde{\alpha_R k_F})^2 + \widetilde{\beta_{SO}}^2}, \quad (S28)$$

and Eq. (S27-2) has a simplified form as shown in Eq. (3) of the main text:

$$ln\left(\frac{T_c}{T}\right) =$$

$$\frac{1}{2}\left[1 - \frac{2(\widetilde{\alpha_R k_F})^2 + \widetilde{\beta_{SO}}^2 - \mu_B^2 B^2}{\rho_+^2 - \rho_-^2}\right] Re\left[\psi\left(\frac{1}{2} + \frac{i\rho_+}{2\pi k_B T}\right) - \psi\left(\frac{1}{2}\right)\right] +$$

$$\frac{1}{2}\left[1 + \frac{2(\widetilde{\alpha_R k_F})^2 + \widetilde{\beta_{SO}}^2 - \mu_B^2 B^2}{\rho_+^2 - \rho_-^2}\right] Re\left[\psi\left(\frac{1}{2} + \frac{i\rho_-}{2\pi k_B T}\right) - \psi\left(\frac{1}{2}\right)\right]. \quad (S27\text{-}3)$$

The result in Eq. (S27-3) is consistent with previous result obtained under clean limit [21]. By setting the Zeeman-type SOI $\widetilde{\beta_{SO}} = 0$, the formula in equation (S27-3) of the supplementary information returns to the case of the in-plane critical field for pure Rashba-type SOI system:

$$ln\left(\frac{T_c}{T}\right) = \frac{1}{2}\left[1 - \frac{2(\widetilde{\alpha_R k_F})^2 - \mu_B^2 B^2}{\rho_+^2 - \rho_-^2}\right] Re\left[\psi\left(\frac{1}{2} + \frac{i\rho_+}{2\pi k_B T}\right) - \psi\left(\frac{1}{2}\right)\right]$$

$$+ \frac{1}{2}\left[1 + \frac{2(\widetilde{\alpha_R k_F})^2 - \mu_B^2 B^2}{\rho_+^2 - \rho_-^2}\right] Re\left[\psi\left(\frac{1}{2} + \frac{i\rho_-}{2\pi k_B T}\right) - \psi\left(\frac{1}{2}\right)\right]$$

with $2\rho_{\pm} \equiv \sqrt{(\mu_B B + \widetilde{\alpha_R k_F})^2 + (\widetilde{\alpha_R k_F})^2} \pm \sqrt{(\mu_B B - \widetilde{\alpha_R k_F})^2 + (\widetilde{\alpha_R k_F})^2}$. Based on numerical calculation, one can easily check that the in-plane critical field for infinite large $\widetilde{\alpha_R k_F}$ satisfies the relation $\frac{\mu_B B_c}{k_B T_c} \leq 1.535$. Thus, in Rashba-type SOI system without spin-orbit scattering, one can obtain the result that the critical field satisfies $B_c < \sqrt{2} B_p$ based on the equations $\mu_B B_P = \frac{\Delta_0}{\sqrt{2}}$ and $2\Delta_0 = 3.53 \times k_B T_c$ (this inequality also holds for our case with $2\Delta_0 = 4.4 \times k_B T_c$), although the spin-orbit scattering can suppress this constraint as shown by the Klemm-Luther-Beasely (KLB) formula.

## VI. Comparison between Zeeman-type SOI mechanism and GL formula

Based on the out-of-plane $B_{c2}$ shown in Fig. S2, we can obtain the real coherence length $\xi_{GL}$ by relation $B_{c2} = \frac{\Phi_0}{2\pi\xi_{GL}^2}(1 - T/T_c)$. And the real coherence length $\xi_{GL} \approx 12.7$ nm for 4ML Pb thin film. The clean limit coherence length for the 4ML Pb thin film can be estimated by the relation $\xi_0 = \frac{\hbar v_F}{\pi\Delta(0)} = \frac{\hbar v_F}{2.2\times\pi k_B T_c} \approx 40$ nm. Here we choose $T_c$=5.78 K and assume relatively small $v_F = 2.5 \times 10^5$ m/s (We want to clarify that $\xi_0$ might be larger than the estimated value if $v_F$ is larger). The mean free path can be estimated by relation $\xi_{GL} \approx \sqrt{\xi_0 \times l_0}$ with $l_0 \approx 4.0$ nm, and the mean free path is much smaller than the coherence length $\xi_0$, which indicates the system locates in the dirty limit. The microscopic theory we developed for Zeeman-protected superconductivity (equation (1) in the main text or equation (S17) in the supplementary information) takes spin-independent scattering into account and can be used in the dirty limit. In Fig. S6(a), the temperature dependence of the Zeeman-protected superconductivity mechanism (the black line) is similar to that of 2D GL formula (the red line) when $T > 0.8T_c$, since equation (1) is proportional to $\sqrt{1 - T/T_c}$ near $T_c$. But the parameters of these two formulas are different. Due to the small thickness of the Pb film, we only consider the theoretical model with Zeeman energy induced by magnetic field and neglect the orbital effect. Furthermore, the in-plane critical field at lower temperatures is important to reveal the difference between the fitting results by these two formulas. In Fig. S6(b)-(d), although the measured in-plane critical field of 4 ML Pb films near $T_c$ can be fitted by the GL formula, the high field data in 6 ML Pb film deviates from the GL formula but can be well fitted by the Zeeman-protected superconductivity mechanism in the relatively low temperature region.

## V. Figures and Tables

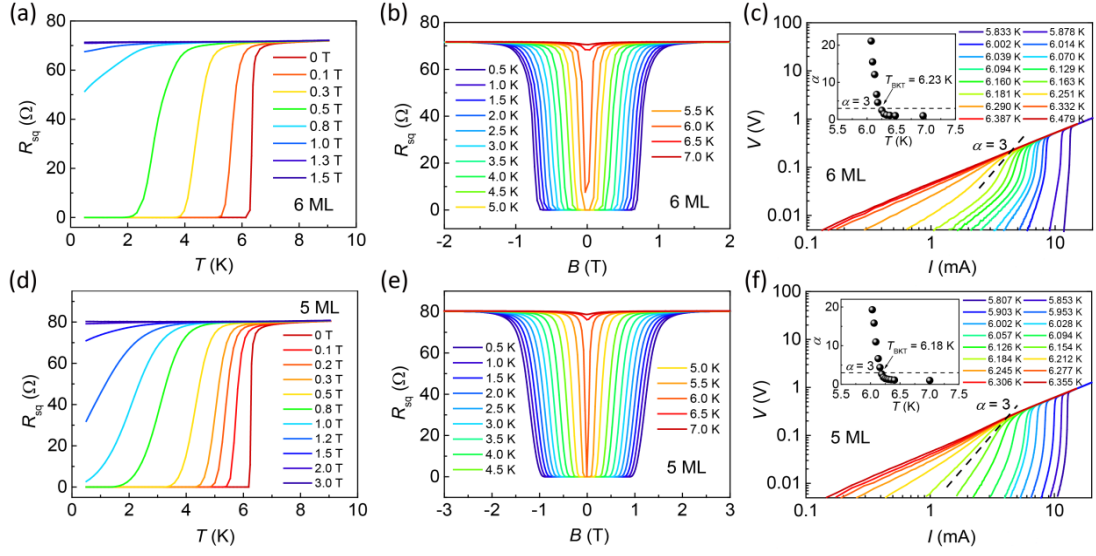

Fig. S1. Transport properties of ultrathin Pb films with thickness of 6 ML and 5 ML under perpendicular magnetic field. (a) (d) Temperature dependence of sheet resistance $R_{sq}$ at various magnetic fields. (b) (e) $R_{sq}(B)$ measured at different temperatures ranging from 0.5 K to 7.0 K. (c) (f) $V(I)$ characteristics in the absence of magnetic field at various temperatures plotted on a double-logarithmic scale. The black line corresponds to $V \propto I^3$ dependence. Inset: power-law exponent $\alpha$ ($V \propto I^\alpha$) as a function of temperature.

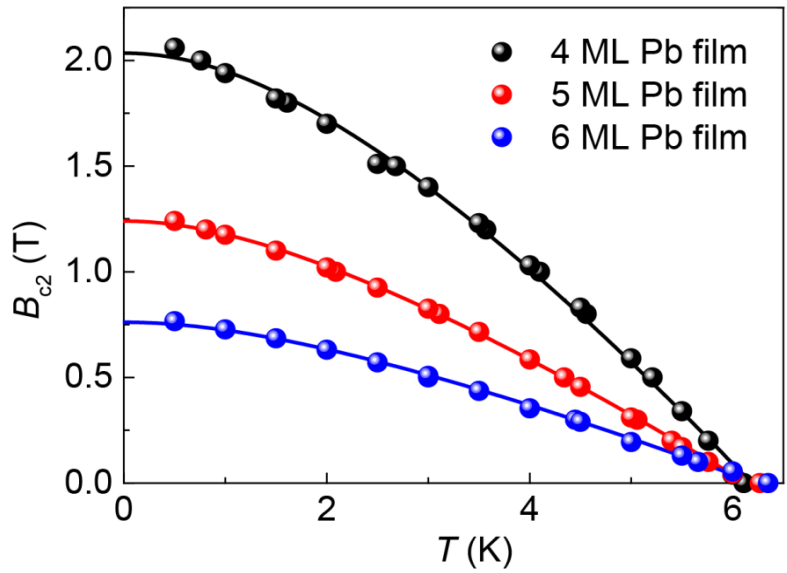

Fig. S2. Perpendicular upper critical field $B_{c2}$ as a function of temperature for 4-6 ML Pb films. Here, $B_{c2}$ is determined as the magnetic field required to reach 50% of $R_n$. The solid lines are fitting curves using Werthamer-Helfand-Hohenberg theory [6], giving larger $B_{c2}(0)$ for thinner films.

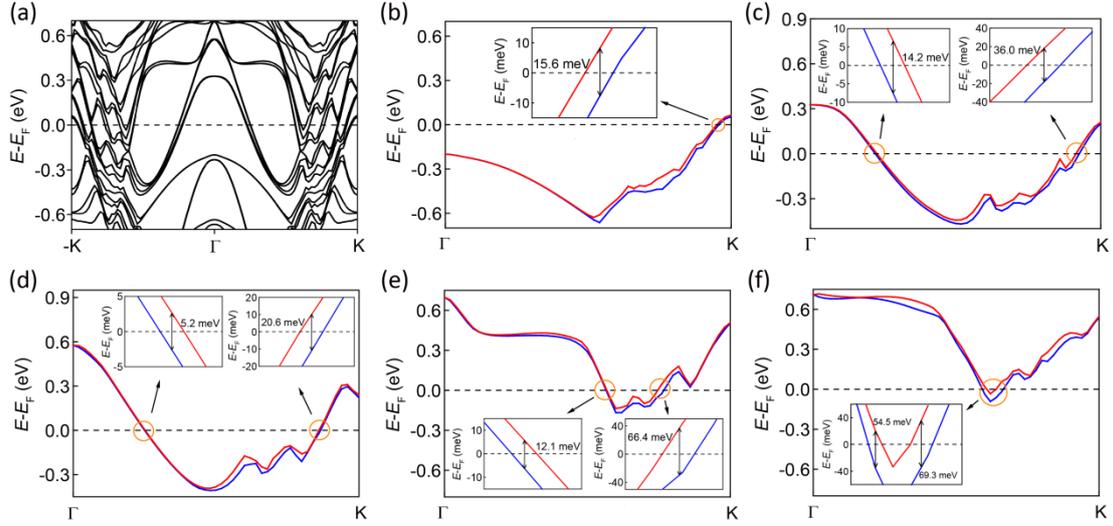

Fig. S3. Calculated band structure of 4 ML Pb film on SIC phase on Si (111) substrate. (a) Band structure of 4 ML Pb film with SOI. (b)-(f) Energy bands along Γ-K that cross the Fermi energy. Red and blue lines represent two bands that are split by SOI. Inset: Energy bands splitting near the Fermi energy due to SOI.

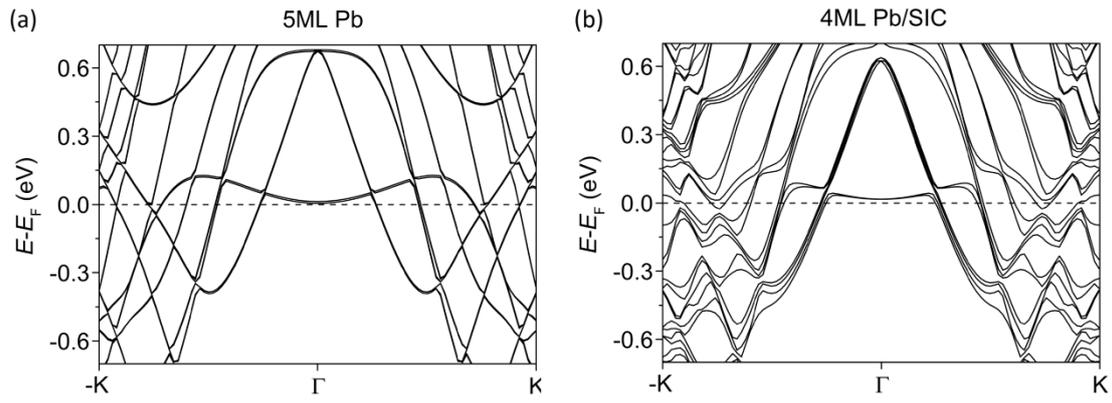

Fig. S4 Calculated band structure of (a) 5ML Pb film, (b) 4ML Pb/SIC system. The spin splitting in 5 ML Pb film is much smaller than that in 4 ML Pb/SIC system, indicating the lattice distortion can induce pronounced spin splitting.

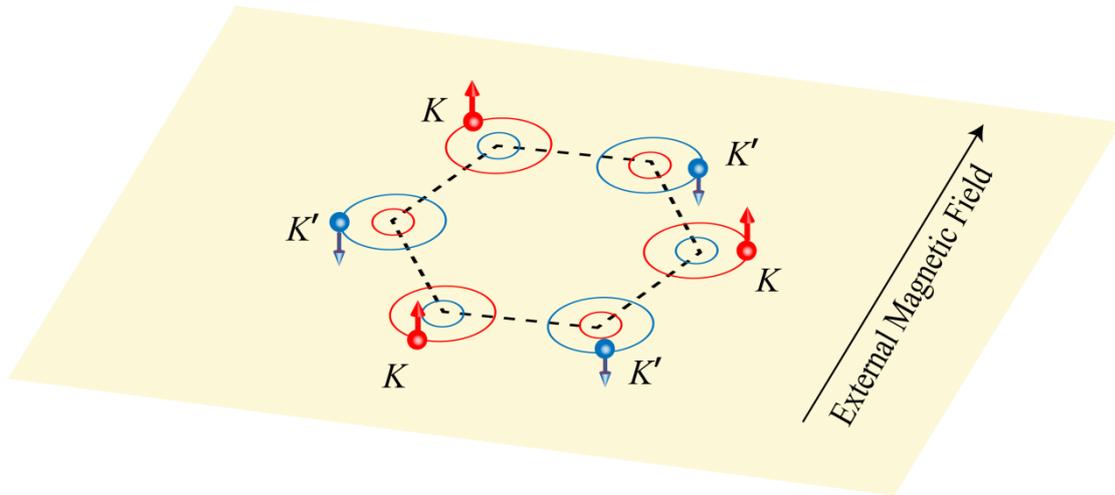

Fig. S5 The spins of the electrons at opposite momentum (k, -k) (which form the Cooper pair) are polarized by Zeeman-type SOI in the opposite out-of-plane orientation.

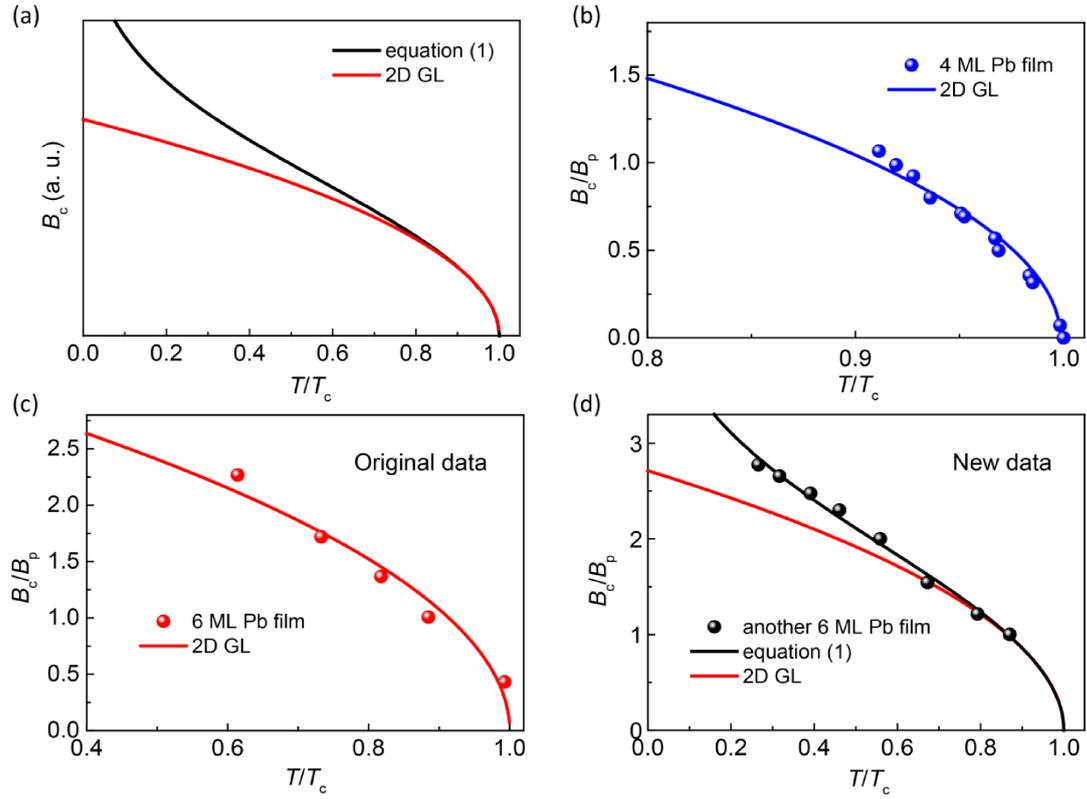

Fig. S6 (a) Theoretical curve of the in-plane critical field $B_c$ of Zeeman-protected superconductivity (based on equation (1) in the main text) (black) and 2D GL theory (red). The two lines are close to each other when $T > 0.8T_c$ but deviate at lower temperatures. (b), (c) 2D GL fitting of the measured $B_c$ in 4 ML and 6 ML Pb film, respectively. (d) the in-plane critical field of another 6 ML Pb film measured in pulsed high magnetic field fitted using equation (1) (black) and 2D GL theory (red).

| Layer | $D_{13}$ /Å | $D_{12}$/Å | $D_{13}/D_{12}$ |
|---|---|---|---|
| SIC | 3.608 | 3.234 | 1.116 |
| 1 | 3.466 | 3.366 | 1.030 |
| 2 | 3.422 | 3.412 | 1.003 |
| 3 | 3.421 | 3.411 | 1.003 |
| 4 | 3.433 | 3.399 | 1.010 |

Table S1. Calculated crystal distortion of 4 ML Pb film on SIC phase on Si (111) substrate. The definition of Pb1, Pb2 and Pb3 is presented in Fig. 3(e). $D_{13}$ is the distance between Pb1 and Pb3 and $D_{12}$ is the distance between Pb1 and Pb2. The ratio of $D_{13}/D_{12}$ shows pronounced lattice distortion in SIC and the upper Pb layers.